%Paper: hep-ph/9308241
%From: FALK@SLACVM.SLAC.Stanford.EDU
%Date: 06 Aug 1993 15:17 -0800 (PST)

\input phyzzx
%
%%%%%%%%%%%%%%%%%%%%%%%%%%%%%%%%%%%%%%%%%%%%%%%%%%%%%%%%%%%%%%%%%%%%%%
%    To obtain a hard copy of this paper, including all figures and
%      tables, send e-mail to  TECHPUB@SLACVM.SLAC.STANFORD.EDU
%       and ask for SLAC-PUB-6311
%%%%%%%%%%%%%%%%%%%%%%%%%%%%%%%%%%%%%%%%%%%%%%%%%%%%%%%%%%%%%%%%%%%%%%
%
%
\def\lam{{\Lambda_{\rm QCD}}}
\def\ol#1{\vbox{\ialign{##\crcr
     \kern1pt\hrulefill\kern1pt\crcr
     \noalign{\kern 1.4pt\nointerlineskip}
     $\hfil\displaystyle{#1}\hfil$\crcr}}}
\def\mev{{\;\rm MeV}}
\def\gev{{\;\rm GeV}}
\def\upket{{|\uparrow\,\rangle}}
\def\dnket{{|\downarrow\,\rangle}}

\def\frac#1#2{\textstyle{#1\over#2}}

\def\pairof#1{#1^+ #1^-}

\def\ee{\pairof{e}}
\def\eea{$\ee$ annihilation}

\def\p{\prime}
\def\pt{{\prime 3}}

\Pubnum{SLAC--PUB--6311\cr JHU--TIPAC-930019}
\date{August, 1993}
\pubtype{T/E}
\titlepage
\title{Production, Decay, and Polarization of Excited Heavy
Hadrons\doeack}
\author{Adam F.~Falk\footnote{\dagger}{On leave from The Johns Hopkins
University, Baltimore, Maryland}\footnote{\ddagger}{Address after
October 1, 1993: Department of Physics 0319, University of California
at San Diego, La Jolla, California 92039-0319} and Michael E.~Peskin}
\SLAC
\abstract{
We discuss the production via fragmentation of excited heavy mesons and
baryons, and their subsequent decay.  In particular, we consider the
question of whether a net polarization of the initial heavy quark may
be detected, either in a polarization of the final ground state or in
anisotropies in the decay products of the excited hadron.  The result
hinges in part on a nonperturbative parameter which measures the net
transverse alignment of the light degrees of freedom in the
fragmentation process.  We use existing data on charmed mesons to
extract this quantity for certain excited mesons. Using this result, we
estimate the polarization retention of charm and bottom baryons.}
\submit{Physical Review D}
\endpage

%\singlespace
%\draft

\chapter{Introduction}

It is well known that many properties of a hadron containing a single
heavy quark $Q$ simplify considerably in the large mass limit
$m_Q\to\infty$.\Ref\intro{E.V.~Shuryak, \sl Phys.~Lett.\bf B93 \rm 134
(1980); W.E.~Caswell and G.P.~Lepage, \sl Phys.~Lett. \bf B167 \rm 437
(1986); E.~Eichten,  \sl Nucl.~Phys. \bf B4  \rm  (Proc.~Suppl.) \rm
170 (1988); G.P.~Lepage and B.A.~Thacker,    \sl Nucl.~Phys. \bf B4 \rm
(Proc.~Suppl.) \rm 199 (1988); N.~Isgur and M.B.~Wise, \sl Phys.~Lett.
\bf B232 \rm  113 (1989).} For $m_Q\gg\lam$, the light degrees of
freedom become insensitive to the mass $m_Q$, and as far as they are
concerned the heavy quark acts simply as a non-recoiling source of
color.  Hyperfine effects associated with the heavy quark
chromomagnetic moment also decouple, and a new ``heavy quark
spin-flavor symmetry'' emerges. There is now an extensive literature in
which this symmetry has been used to make rigorous, model-independent
predictions relating heavy hadron spectra, weak matrix elements and
strong decay rates. Corrections to the $m_Q\to\infty$ limit, both
radiative and nonperturbative, have been explored in great
detail.\Ref\reviews{Recent reviews of this subject include B.
Grinstein, in {\it High Energy Phenomenology}, R.~Huerta and
M.A.~Perez, eds. (World Scientific, 1992); H.~Georgi, in {\it
Perspectives in the Standard Model} (Proceedings of TASI-91),
R.K.~Ellis, C.T.~Hill and J.D.~Lykken, eds. (World Scientific, 1992);
M.~Neubert, SLAC--PUB--6263 (1993), to appear in Physics Reports.}

In this article we will apply the same symmetries to the production of
heavy mesons and baryons.  In the limit $m_Q\to\infty$ such a process
factorizes into short-distance and long-distance pieces.  A heavy quark
$Q$ is first produced via some high energy interaction, perhaps as part
of a pair $Q\ol Q$ with large relative momentum.  This process, for
example the decay of a virtual photon or $Z$ boson, is typically
calculable in perturbation theory. This perturbative stage is finished
in a time short compared to the time scale of the nonperturbative
strong interactions. Over a longer time scale, a fragmentation process
occurs which eventually forms a physical hadron containing the heavy
quark. One might visualize this process as the splitting of a color
flux tube which joins the heavy $Q$ to the other colored products of
the hard reaction. However one models the fragmentation process, it
occurs entirely at  length scales of order $1/\lam$, and hence involves
the redistribution of energies small compared to $m_Q$.  As a result,
the velocity of $Q$ remains unchanged once it has been produced, and
its mass and spin, which are determined by the calculable
short-distance physics, decouple from the nonperturbative dynamics.
The situation here is entirely analogous to that of the much-explored
weak decays of heavy hadrons.  This analogy has already been exploited
in discussions of the production of ground state pseudoscalar and
vector mesons.\Ref\oldanal{A.F.~Falk and B.~Grinstein, \sl Phys.~Lett.
\bf B249 \rm 314 (1990); T.~Mannel, W.~Roberts and Z.~Ryzak, \sl
Phys.~Lett. \bf B247 \rm 412 (1990); T.D.~Cohen and J.~Milana, \sl
Phys. Lett. \bf B306, \rm 134 (1993).}

It is tempting to generalize this philosophy directly to the production
of excited heavy mesons and heavy baryons.  For these systems, a major
issue is the question of the polarization of the heavy state along the
axis of fragmentation.  We will show that when one computes this
polarization, the factorization of heavy- and light-quark physics in
the fragmentation process is not quite so straightforward.

Two new ingredients enter the analysis.  First, it is often the case
that the strict heavy quark approximation fails for the last stage of
fragmentation in systems with $c$ and $b$ quarks.  We will present some
examples in which light-quark rearrangements, with rates formally
independent of $m_Q$, are slowed by phase space or angular momentum
factors so that they become comparable to the rate, of order
$(m_Q)^{-1}$,  for  processes that flip the heavy quark spin.

The possibility to transfer angular momentum from the heavy to the
light degrees of freedom means that the final heavy quark polarization
will depend on the polarization of the light degrees of freedom created
in the fragmentation process. This brings in the second new feature of
the analysis.  Since fragmentation is a strong interaction process
which conserves parity, it cannot select a prefered spin direction
along the axis of fragmentation.  However, the strong interactions can
produce the light degrees of freedom in a way which is anisotropic
about this axis, for example, preferring states with longitudinal to
those with purely transverse polarization.  We will define parameters
$w_j$ which characterize the alignment of light degrees of  freedom of
spin $j$ and show how these affect the polarization of the heavy
hadrons and their decay products.  The $w_j$ are new parameters of
potential importance which provide nontrivial tests of fragmentation
models.

Our analysis is organized as follows:  In Section 2, we will
give a more detailed discussion of the relative time scales in heavy
quark fragmentation.  In Section 3, we will discuss the polarization of
heavy quarks in ground state heavy mesons $D$, $D^*$ and $B$, $B^*$.
This is the simplest case,  but we will see that here all polarization
information is lost in the fragmentation process. In Section 4, we will
discuss the polarization of excited heavy  mesons. Here we will
identify reactions in which light-quark processes are hindered below
the heavy quark spin flip time.  This affects the dependence of the
heavy meson decay distributions on the fragmentation orientation.  We
will determine the orientation parameter $w_{3/2}$ from data on excited
charmed mesons.

\REF\MS{T. Mannel and G. A. Schuler, \sl Phys. Lett. \bf B279, \rm
    194 (1992).}
\REF\CLKPS{F. E. Close, J. K\"orner, R. J. N. Phillips, and D. J.
    Summers, \sl J. Phys. \bf G 18, \rm  1716 (1992).}
In   Section 5, we will discuss the polarization of heavy baryons. The
ground state heavy baryon is the $\Lambda_Q$, the bound state of a
heavy quark with a light diquark system of spin 0.  Using this
identification and the $m_Q\to \infty$  limit, Mannel and Schuler
\refmark{\MS}  and Close, K\"orner, Phillips, and Summers
\refmark{\CLKPS} have argued that $\Lambda_b$'s produced at the $Z^0$
resonance should be highly polarized.  The second of these groups also
pointed out the potential for depolarization when $\Lambda_b$'s are
produced by the decay of excited baryons $\Sigma_b$ and $\Sigma_b^*$.
We will discuss this effect quantitatively and show that it potentially
leads to significant depolarizations in an interesting pattern.  These
effects can also be seen in the  study of charmed baryons.  We will
show how these effects are sensitive to the basic parameters governing
baryon  fragmentation and decay and suggest ways to determine these
parameters experimentally.

\chapter{Time Scales in Heavy Quark Fragmentation}

We are concerned in this paper with the dynamics of the spin of a heavy
quark produced in a fragmentation process.  To begin, we will discuss
in this section the various time scales which arise in heavy quark
fragmentation.  This will provide a consistent framework for our later
analysis.

We always imagine that we begin with a heavy quark which has been
ejected at relativistic speed from a hard reaction.   We will compute
time in the frame of the heavy quark.  The axis linking this frame to
the center-of-mass frame of the hard process is a preferred direction,
which we call the axis of fragmentation.  We will take the $\hat 3$
axis to lie along this line, pointing in the direction of the
heavy-quark velocity.

In the rest frame of the heavy quark, the leading operator which
couples to the heavy-quark spin is the color magnetic moment operator,
whose coefficient is suppressed by $1/m_Q$.  Thus, the rate of heavy
quark spin flip is very slow on the scale of $\lam$.  We might imagine
the early stages of fragmentation  to involve the production of highly
excited mesons or baryons containing the heavy quark, which then
rapidly eject pions and decay to lighter excited states.  Throughout
this process, the heavy quark spin retains its initial orientation. The
process continues until we reach a state whose lifetime is comparable
to the time required to flip the heavy quark spin.

This long-lived heavy quark state is characterized by two angular
momenta: $s = \coeff12$, the heavy quark spin, and $j$, the spin of the
light degrees of freedom.   The combination gives states of total spin
$J = j \pm \coeff12$, which we will call $H$ and $H^*$.  The color
magnetic moment interaction produces a small mass splitting between $H$
and $H^*$ which we call $\Delta$.  This energy splitting $\Delta$ can
be identified with the rate of heavy quark spin flip processes in the
$(H,H^*)$ multiplet.

The states of the heavy quark multiplet can decay either by transitions
involving the heavy or light quarks separately or by transitions $H^*
\to H$.  In the former case, $H$ and $H^*$ have the same decay rate,
$\Gamma$.  We will call the rate of the $H^* \to  H$ transition
$\gamma$.  This latter decay is a QCD or QED magnetic dipole
transition.  Thus, it is suppressed by two powers of $1/m_Q$ from the
square of the matrix element and by further powers from the phase
space.  We expect, then, that $\gamma\ll\Delta$.  On the other hand,
the overall decay rate $\Gamma$ may have an arbitrary relation to these
two parameters.

To visualize the roles of the three rates $\Delta$, $\Gamma$, and
$\gamma$, it is useful to think about the three possible extreme cases:

1. $\Gamma\gg \Delta \gg\gamma$:  In this case, the heavy hadrons
decay so rapidly that the color magnetic moment interactions of the
heavy quark with the light degrees of freedom do not have time to work.
If $\Gamma$ is a rate of a strong interaction decay process, then in
this case the multiplet $(H, H^*)$ would belong to the early stages of
fragmentation, in the sense described above, and transitions through
this multiplet would have no effect on the heavy quark spin dynamics.
Another possibility, if the quark mass is extremely large, is that the
dominant contribution to $\Gamma$ could come from the heavy quark weak
decay.   In this circumstance, as long as $\Gamma \gg \Delta$, the weak
interaction decay will measure a spin orientation for the heavy quark
which is the same as that which was produced in the hard process, with
no depolarization  by fragmentation.  This is the case which typically
arises in studies of the top quark.\Ref\toprefs{I. Bigi and H.
Krasemann, \sl Z. Phys. \bf C7, \rm 127 (1981); J. K\"uhn, \sl Acta
Phys. Austr. \rm Suppl. \bf XXIV, \rm 203 (1982); I. Bigi, Y.
Dokshitzer, V. Khoze, J. K\"uhn, and P. Zerwas,  \sl Phys. Lett. \bf
B181, \rm 157 (1986).} Notice that the approximation $\Gamma \gg\Delta$
can be valid even if $\Gamma \sim \lam$, so that the heavy quark
partially hadronizes before it decays.

2. $\Delta \gg \Gamma \gg\gamma$:  In this case, the heavy hadron
states $H$ and $H^*$ form distinct resonances.  These resonances have
width $\Gamma$ and are well separated from one another. The decay
products reflect the heavy quark spin orientation in the separate
states $H$ and $H^*$.  These two contributions must be added
incoherently; thus, the heavy quark is depolarized from its initial
orientation.  In Sections 4 and 5, we will given examples in which this
limit applies even though $\Gamma$ is the rate of a strong interaction
decay process.

3. $\Delta\gg \gamma\gg \Gamma$:  In this case, the heavy hadrons $H^*$
have time to make the transition to $H$ before undergoing a decay out
of the multiplet.  In this case, the decay products of the multiplet
reflect only the heavy quark spin orientation in the state $H$. This
leads to a substantial (and sometimes complete) depolarization. The
simplest example of this situation arises in the production of $B$ and
$B^*$ mesons in fragmentation; we will discuss this example in Section
3.

In our arguments in the next few sections, we will begin by assuming
that the initial heavy quarks produced by the hard process are
completely polarized.  At some stage, though, we must go over to the
realistic situation in which they are produced with partial
polarization. We will denote the initial heavy quark polarization by
$P$. Since the $Z^0$ resonance provides the most accessible source of
polarized heavy quarks, and since $Z^0$ decays produce mainly
left-handed quarks, we will define the polarization to be positive in
this case. At the $Z^0$,
$$P_q = A_{LR}^q =
  {g_{Lq}^2 - g_{Rq}^2 \over g_{Lq}^2 + g_{Rq}^2}, \eqn\Pdeterm$$
so that
$$P_b = 0.94 , \qquad     P_c = 0.67 \  ,   \eqn\Porig$$
for $\sin^2\theta_{w*} = 0.232$.
In the course of this paper, we will investigate what fractions of
these very large values are actually visible to experimenters.

\chapter{Heavy Pseudoscalar and Vector Mesons}

The simplest example with which to start is that in which the light
degrees of freedom have spin-parity $j^P=\coeff12^-$.  The constituent
quark model would suggest that such a state, consisting of a light
antiquark in an $S$-wave, is the one of lowest energy, and in the charm
and bottom systems this has indeed been observed to be the case.  This
light quark system combines with the heavy quark $Q$ to form the
multiplet $(H, H^*)$ consisting of a heavy pseudoscalar meson and a
heavy vector meson. The states are split by an amount of order
$\lam^2/m_Q$.  In the charm system, this is the $(D, D^*)$  multiplet;
for bottom, it is the $(\ol B, \ol B^*)$ system.    In the following
discussion, we will refer to the spin of the light degrees of the
freedom loosely as the `spin of the antiquark'.

In the charm case, most of the parameters of this system are well
determined. The $D$--$D^*$ splitting $\Delta$ is approximately
$140\mev$.  Although $\Delta>m_\pi$ and the strong decay $D^*\to D\pi$
occurs, it is so suppressed by phase space that as yet there is only an
upper limit on the intradoublet transition width, $\gamma <1.1\mev$ for
$D^{*0}$,  $< 2\mev$ for $D^{*+}$. However, quark model estimates lead
one to believe that $\gamma$ should be no more than an order of
magnitude smaller than this upper bound.  Finally, since the $D$ meson
can only decay weakly, its width $\Gamma$ is extremely small, of the
order of $10^{-10}\mev$.  Hence we are safely within the region
$\Delta\gg\gamma \gg\Gamma$ discussed above.   A similar picture
applies for the bottom mesons. Here $\Delta = 46\mev$. Because the
strong decay $\ol B^*\to\ol B\pi$ is prohibited, the transition must
occur electromagnetically.  The width $\gamma$ for $\ol B^*\to B\gamma$
may be estimated from the upper limit on $D^*\to D\pi$ and the
branching ratio for $D^*\to D\gamma$; we find an approximate value
$\gamma \sim 0.01\mev$. The multiplet width $\Gamma$ is again due to a
weak decay and so is many orders of magnitude smaller.  In both cases,
we are in the situation of case 3 described in Section 2.  For
concreteness, we will refer to the bottom system in the following
discussion.

We begin with the case in which  the  initial $b$ quark is completely
polarized in the left-handed direction.  We would like to investigate
whether any information on the initial $b$ polarization can be
recovered experimentally.  The fragmentation process leads to a heavy
meson in which the $b$ is combined with an antiquark (more carefully,
with light degrees of freedom with $j = \coeff 12$). We may assume that
the fragmentation process occurs so rapidly that the color magnetic
forces do not have time to act; thus the spin of the antiquark is
uncorrelated from the spin of the $b$.  In this case, there are only
two choices for the spin orientation of the antiquark: $j^3 = \pm
\coeff12$; we must sum over these possibilities incoherently. Since the
fragmentation process conserves parity, the antiquark spin cannot be
preferentially aligned in one direction along the axis of
fragmentation; thus, the two choices occur with equal probability.
Hence, the result of fragmentation is to produce meson states with the
quark and antiquark spins
$$\dnket_b\dnket_{\bar q},\qquad
  \dnket_b\upket_{\bar q} \eqn\bothkets$$
with equal probability. Notice that the second state in \bothkets\ is a
linear combination of a $\ol B$ and a $\ol B^*$ meson.  The two
components of this state propagate coherently up to a time
$\Delta^{-1}$ and then go out of phase with one another.  Since, in
this example, $\Delta \gg \gamma \gg \Gamma$, the $\ol B$ and $\ol B^*$
components become completely incoherent before any decay occurs.  This
gives rise to the following table of probabilities for the occupation
of the various possible helicity states:
$$\pmatrix{ p(\ol B^*, h )\cr p(\ol B,h)\cr } \ = \
  \pmatrix{ \coeff12 & \coeff14 & 0 \cr & \coeff14 & \cr} .
  \eqn\bhelic$$
The helicity of the $\ol B$ runs across the table from negative to
positive values; for example, the table assigns the state $\ol
B^*(h=-1)$ the probability $\coeff 12$.

At a time $\gamma^{-1}$, the $\ol B^*$ mesons decay electromagnetically
to $\ol B$'s.   After this point, the $\ol B$'s will contain no
polarization information, since the $\ol B$ meson has spin zero.  Thus,
the polarization information can only be encoded in the photons emitted
in the decay.

The decay $\ol B^*\to\ol B\gamma$ proceeds primarily through the light
quark magnetic moment operator
$${e_q{\bf\sigma}_q\cdot{\bf B}\over2m_q}\,,\eqn\magop$$
since the $b$ magnetic moment is suppressed by $1/m_b$.  Let $\theta$
be the angle between the photon momentum and the fragmentation axis, in
the $\ol B^*$ rest frame.  Then the differential partial widths
$d\gamma/d\cos\theta$ for the various $\ol B^*$ helicity states are
proportional to
$$\eqalign{\ol B^*(\pm1)&:\cr \ol B^*(0)&:\cr}\qquad
  \eqalign{&\textstyle{\coeff12}(1+\cos^2\theta)\,,\cr
  &\sin^2\theta\,.\cr}\eqn\bstardis$$
Multiplying these rates by the probabilities for producing the helicity
states $\ol B^*(\pm1)$ and $\ol B^*(0)$, we find that the total
distribution is proportional to
$$\coeff14(1+\cos^2\theta)+ \coeff14\sin^2\theta=\coeff12
  \eqn\firstnowin$$
Hence, the photons are emitted isotropically, and their angular
distribution gives no polarization information.  The emitted photons
are preferentially polarized left-handed, but this polarization cannot
be observed by a standard high-energy particle detector.  We conclude
that the polarization of the $b$ quark is unobservable in fragmentation
to $\ol B$ and $\ol B^*$ mesons.

This is our first example of a `no-win' theorem, to which we shall
return.  Under most conditions, the angular distribution of decay
products gives no information on the polarization of the heavy quark.
In reaching this conclusion, we do not assume that the heavy quark spin
is decoupled from the decay process.  In this example, the heavy quark
spin couples to the light antiquark, giving it a net polarization
$\coeff 12$ on a time scale of order $\Delta^{-1}$.   However, the
strong and electromagnetic interactions responsible for the decay
conserve parity and thus cannot be sensitive to the direction of the
heavy quark spin.  Thus, the angular distribution of the decay products
is the same as as it would be if we averaged over the two possible
directions of the heavy quark spin. There is one amusing exception to
this rule, which we will discuss in Section 4.

\REF\CLEOpol{Y. Kubota, \etal  (CLEO Collaboration),
    \sl Phys. Rev. \bf D44, \rm 593 (1991).}
\REF\PDG{K. Hikasa, \etal\
   (Particle Data Group), \sl Phys. Rev. \bf D45 \rm (1992), No. 11-II.}
\REF\CLEODSTARD{F. Butler, \etal\ (CLEO Collaboration), \sl
    Phys. Rev. Lett. \bf 69, \rm 2041 (1992).}
As a footnote to this section, we comment on the validity of the
helicity distributions \bhelic \ for the charmed mesons. The heavy
quark limit predicts that, when we average over the direction of the
heavy quark spin, we recover the naive spin-counting predictions that
the  $D$ and $D^*$ mesons are produced in a 1:3 ratio, and that the
$D^*$ mesons are unpolarized.  The latter result is confirmed by a CLEO
measurement\refmark\CLEOpol\ which finds only a few percent
longitudinal polarization in $D^*$'s produced directly from \eea.
However, many groups have measured the ratio $P_V = (D^*)/(D + D^*)$,
which spin-counting predicts to be 0.75, and find a substantially
smaller number:\Ref\Oursumm{This value of $P_V$ is found by combining
the following experimental results:  D. Bortoletto, \etal\ (CLEO
Collaboration), \sl Phys. Rev. \bf D37, \rm 1719 (1988); H. Albrecht,
\etal\ (ARGUS Collaboration), \sl Z. Phys. \bf C52, \rm 353 (1991); D.
Decamp, \etal \ (ALEPH Collaboration), \sl Phys. Lett. \bf B266, \rm
218 (1991);  F. Hinode, \etal\ (VENUS Collaboration), KEK preprint
92-192 91993).  The values of $P_V$ given in these papers must be
corrected to a common set of branching ratios:  $B(D^0 \to K^-\pi^+) =
3.65 \pm .21$ \% (ref. \PDG); $B(D^{*+} \to D^0 \pi^+) = 68.1 \pm 1.6$
\% (ref. \CLEODSTARD); then the various determinations are in good
agreement.  We thank Sheldon Stone for discussions of these results.}
$$P_V = 0.65 \pm .06   \ .\eqn\valueofPV$$
Such a value would not be unexpected in a thermodynamic model of
particle production in which the higher-mass states are suppressed by a
factor
$$\exp\bigl[ - \Delta m/ T_H \bigr] \ ,\eqn\thebigTfactor$$
where $\Delta m$ is the $D^*$--$D$ mass difference and $T_H$ is a
hadronic `temperature', which should be expected to be about 300 MeV.
Indeed, the central value of \valueofPV is reproduced by setting
$T_H = 280$ MeV.
Notice that the suppression factor \thebigTfactor\ does formally tend
to 1 in the heavy quark limit in which members of the same heavy-quark
multiplet become degenerate.  However, for the charmed mesons,  it gives
almost a factor 2 suppression.
  The correction results from the fact that
the excited charm states which decay to $D$ and $D^*$ have widths which
are comparable to the $D^*$--$D$ mass difference and so can resolve
these two states and prefer the lighter $D$. This is a first example of
the competition between decay rates and mass splittings which we will
discuss quantitatively in  the later sections of this paper.

In the examples discussed later in this paper, we will continue to
ignore the thermodynamic factor \thebigTfactor\ in the initial
probability distributions of heavy mesons. In those later examples,
this assumption will be justified by the fact that the states which
decay to the $(H,H^*)$ multiplet in those cases typically have widths
much larger than the $H$--$H^*$ mass splitting.

\chapter{Excited Heavy Mesons}

We now turn to the more complicated case of heavy mesons in which the
light degrees of freedom are in an excited state.  We will focus on the
charm system, and in particular on the observed excited charmed mesons
$D_1(2420)$ and $D_2^*(2460)$.  We will discuss the decay distributions
of these states from the viewpoint of heavy quark symmetry.

In the quark model, the lowest-energy excited states of the $D$ and
$D^*$ mesons should be states in which the light antiquark has one unit
of orbital angular momentum.  By coupling this angular momentum to the
antiquark spin, we find states in which the light degrees of freedom
have $j^P = \coeff 12^+$ and $\coeff32^+$. In the $m_c\to \infty$
limit, the angular momentum $j$ is a good quantum number irrespective
of its quark model interpretation.

\REF\IW{N.~Isgur and M.B.~Wise,\sl Phys.~Rev.~Lett. \bf 66 \rm 1130
    (1991).}
\REF\FL{A. F.~Falk and M.~Luke,\sl Phys.~Lett. \bf B292  \rm 119
    (1992).}
It is reasonable to identify the  spin-1 $D_1(2420)$ and the spin-2
$D_2^*(2460)$ as the heavy meson multiplet $(H,H^*)$  with $j^P =
\coeff32^+$.\refmark{\IW, \FL} The $j^P=\coeff12^+$ doublet, consisting
of a spin-0 ($D_0^*$) and a spin-1 ($D_1'$) meson, has not yet been
identified.  At order $1/m_c$, there may be mixing between the $D_1$
and the $D_1'$ states, since they have identical quantum numbers.

It is likely that the $(D_0^*,D_1')$ doublet has not been found because
these states have a very large decay width to $D$ and $D^*$.  They
should decay by emitting a pion in the $S$-wave, a completely open
channel. Kaidalov and Nogteva\Ref\kaidalov{A.B.~Kaidalov and
A.V.~Nogteva, Sov.~J.~Nucl.~Phys. \bf 47, \rm 321 (1988).} have
estimated the width $\Gamma$ for this multiplet to be several hundred
MeV.  On the other hand, the mass splitting $\Delta$ should be smaller
than 40 MeV, the mass splitting of the $j= \coeff32$ multiplet.  Thus,
this doublet corresponds to the uninteresting case 1 of Section 2,
$\Gamma \gg \Delta$.

\REF\argus{H.~Albrecht {\it et al.} (ARGUS Collaboration),
    \sl Phys.~Lett.~\bf B221 \rm 422 (1989), \bf B232, \rm 398 (1989).}
\REF\cleo{P.~Avery {\it et al.} (CLEO
     Collaboration), \sl Phys.~Rev. \bf D41 \rm 774 (1990).}
The situation for the observed $D_1$ and $D_2^*$ is more interesting.
Since the $j^P$ of the light degrees of freedom changes from
${\coeff32}^+$ to ${\coeff12}^-$, the decay pion must be emitted into
an orbital $D$-wave, and so the decay width is suppressed by angular
momentum factors. The observed decay width $\Gamma$ of the two members
of the doublet is about 20 MeV, while the observed splitting $\Delta$
is approximately 35  MeV.\refmark{\argus,\cleo} The intradoublet
transition is an electromagnetic  decay, so $\gamma$ is much smaller
than either of these rates.  In the following discussion, we will treat
the decays of $D_1$ and $D_2^*$ in the limit $\Delta \gg \Gamma \gg
\gamma$, case 2 of Section 2.  This is justified as a first
approximation:  Since the $D_1$ and $D_2^*$ peaks are well separated
compared to their width, their decays can be treated incoherently.

Because the experiments of refs. \argus\ and \cleo\ were carried out
well below the $Z^0$, the  charmed quarks were produced from \eea\ with
no polarization.  Nevertheless, for full generality, we will begin our
analysis by assuming that the charmed quarks have complete left-handed
polarization.  To this polarized charmed quark, we must add the light
$j=\coeff 32$ system.  This system can be formed in one of four
possible helicity states.  Parity invariance requires that the
probability of forming a given helicity state  cannot depend on the
sign of this helicity $j^3$.  However, states with different magnitudes
$|j^3|$ can have different probabilities.  For the examples discussed
in this paper, we can characterize these probabilities in the following
way: {\it For a system of light degrees of freedom of spin $j$, let
$w_j$ be the probability that fragmentation leads to a state with the
maximum  value of $|j^3|$.}  The parameter $w_j$ takes values between 0
and 1.

In the case at hand, the various helicity states of the light degrees
of freedom appear with the probabilities
$$ p({\coeff32},j^3) = \bigl( \coeff12 w_{3/2},
   \coeff12 (1-w_{3/2}),
   \coeff12 (1-w_{3/2}),
   \coeff12 w_{3/2} \bigr),
   \eqn\thethalfprobs$$
where the helicity $j^3$ of the light degrees of freedom runs across
the table from $-\coeff32$ to $\coeff32$. The state of definite
left-handed $c$ spin, combined with the state of the light degrees of
freedom of definite $j^3$, produces a coherent linear superposition of
the $D_1$ and $D_2^*$ states of helicity $h = j^3 - \coeff12$. In a
time $\Delta^{-1}$ into the fragmentation process,  the $D_1$ and
$D_2^*$ components of this state become incoherent and it becomes
appropriate to describe the original state as a mixed state containing
$D_1$ or $D_2^*$ with fixed probabilities.  Following this logic, we
find that the possible helicity states of  $D_1$ and $D_2^*$ should be
populated with the probabilities shown in the following table:
$$ \pmatrix{ p(D_2^*, h )\cr p(D_1,h)\cr} \ = \
   \pmatrix{ \coeff12 w_{3/2}& \coeff38(1-w_{3/2})
   & \coeff14 (1-w_{3/2}) & \coeff 18 w_{3/2}& \phantom{w}0\phantom{w}
   \cr & \coeff18(1-w_{3/2}) & \coeff14 (1-w_{3/2})& \coeff 38 w_{3/2}
   & \cr} .
   \eqn\chelic$$
The notation is as in \bhelic, with the values of the helicity running
from negative to positive across the table. To find the probabilities
for charmed quarks with initial polarization $P=0$, average the
probabilities of states with equal and opposite helicities.  Notice
that for any value of $w_{3/2}$, and for any $P$, the total
probabilities for producing the spin-2 and spin-1 states are $\coeff58$
and $\coeff38$, respectively.

Given these probabilities, we may now compute the angular distributions
for the observed decays $(D_1,D_2^*)\to(D,D^*)+ \pi(p)$.  The general
theory of pion transitions between heavy hadrons is due to Isgur and
Wise,\refmark{\IW} and is reviewed in Appendix A.  According to this
theory the rate for the pion transition from a heavy hadron  with light
degrees of freedom with spin $j$ to a heavy hadron with light degrees
of freedom freedom of spin $j'$ depends on the total spins $J$, $J'$ of
the initial and final hadrons according to the factor
$$ (2j+1)(2J' + 1)
   \left| \left\{\matrix{j' & j & L \cr
   J& J' &\coeff 12\cr}\right\} \right|^2 \cdot p_\pi^{2L+1} .
   \eqn\totalratec$$
In this equation, $L$ is the pion orbital angular momentum, the bracket
denotes a 6-$j$ symbol, and $p_\pi$ is the pion 3-momentum.  For the
transitions from $(D_1,D_2^*)$  to $(D,D^*)$, $L=2$. The last factor in
\totalratec\ is the kinematic suppression factor for emitting pions of
large $L$, which may vary significantly over the heavy multiplets  even
if their splitting is small.  The purely group theoretic factors
give\refmark{\IW}
$$ \eqalign{
   \Gamma(D_1\to D\pi):\Gamma (&D_1 \to D^* \pi):
   \Gamma(D_2^*\to D\pi) : \Gamma(D_2^* \to D^* \pi)  \cr
   & = \quad  0  \quad : \quad 1 \quad : \quad
   \coeff25  \quad : \quad \coeff 35\ . \cr}
   \eqn\IWrates$$
The kinematic factor $p_\pi^5$ for these decays are
$$  4.5\ : \ 0.90 \ : \ 6.2 \ : \ 1.4  , \eqn\kinemDDpi$$
in units of $10^{-2}$ GeV$^5$.

We can use these numbers to assess the experimental validity of the
heavy quark approach to $(D_1,D_2^*)$ decays.  Our discussion here
follows the work of  Lu, Wise, and Isgur (LWI).\Ref\LWI{M.-L. Lu, M. B.
Wise, and N. Isgur, \sl Phys. Rev. \bf D45 \rm (1992) 1553.} Assembling
the factors above, one  finds
$$\Gamma(D_2^*\to D\pi)/\Gamma(D_2^* \to D^* \pi) = 3.0,
  \eqn\kinpred$$
independent of charge assignments; this is in good agreement with the
Particle Data Group average  of $2.4 \pm 0.7$ for the relative rates of
$D_2^{*0} \to D^+ \pi^-, D^{*+}\pi^-$.\refmark\PDG\ From these values
and the  observed $D_2^{*0}$ width of $19\pm 7$  MeV, one predicts the
total width of the $D_1$ meson to be $\Gamma(D_1)  =  5 \pm 2$ MeV,
which is substantially smaller than the observed   value of $20 \pm 7$
MeV for the $D_1^0$.   LWI ascribed the discrepancy to a small mixing
of the $D_1$ with the $D_1'$. An increment of the  $D_0$ width by 10
MeV, which would be accomplished by a mixing angle of order 0.2, would
be quite sufficient. Such a mixing angle is not unreasonable, since the
mixing is expected to be of order (300 MeV/$m_c$).  LWI proposed an
experimental test of this idea, which we will return to below.

We now add to these results our understanding of fragmentation to heavy
mesons.  This will allow us to compute the angular distributions of the
$D_1$ and $D_2^*$ decay products in terms of the parameter $w_{3/2}$.
We begin with the decay $D_2^* \to D \pi$. Let $\theta$, $\phi$ denote
the orientation of the pion with respect to the fragmentation axis, as
measured in the $D_2^*$ rest frame.  The amplitude for the production
of a pion at $\theta,\phi$ from a $D_2^*$ meson of helicity $h$ is
proportional to $Y_{2h}(\theta, \phi)$.  Thus, the complete  pion
angular distribution should be proportional to
$$ \sum_h  p(D_2^*,h)  \bigl| Y_{2h}(\theta,\phi) \bigr|^2,
   \eqn\firstYs$$
where $p(D_2^*,h)$  are the probabilities from \chelic.  Expanding and
normalizing, we find
$$ {1\over \Gamma}{d\Gamma\over d\cos\theta}(D_2^* \to D \pi) =
   \coeff14\Bigl[1   +  3 \cos^2\theta - 6 w_{3/2}
   \bigl( \cos^2\theta - \coeff 13\bigr)\Bigr] \ .
   \eqn\distone$$
Note that this distribution is invariant under $\cos\theta \to -
\cos\theta$, as required by parity, and thus gives no information on
the $c$ quark polarization. This accords with the `no-win' theorem
discussed at the end of Section 3. The pion angular distribution is
generally anisotropic but becomes isotropic for isotropic fragmentaion,
$w_{3/2} = \coeff12$.  In fact, the dependence of \distone\ on
$w_{3/2}$ is fixed by this requirement and the requirement that the
total rate be independent of $w_{3/2}$.

This angular distribution has been measured by ARGUS,\refmark{\argus}\
and so the parameter $w_{3/2}$ can be extracted from experiment.
\FIG\FindtheW{Angular distribution of pions in the decay $D_2^*\to
D\pi$.  The data are shown, along with theoretical predictions
corresponding to $w_{3/2}=0$ (solid curve) and $w_{3/2}=0.2$ (dashed
curve).} The ARGUS data are shown in Fig.~\FindtheW, along with the
theoretical predictions for $w_{3/2}=0$ and $0.2$.  The ARGUS analysis
found no significant population of the extreme helicity states $h = \pm
2$.  This implies that $w_{3/2}$ is small.  Our best fit would come at
$w_{3/2}= -0.3$, if this were meaningful.  Assuming that $w_{3/2} > 0$,
we find
$$ w_{3/2} <  0.24, \qquad  90\% \ {\rm conf.}
   \eqn\wthalfbound$$
We will discuss the physical interpretation of this result below.

Once $w_{3/2}$ is known, we have definite predictions for the angular
distributions of the remaining excited $D$ meson decays.  Consider
next the decay $D_2^* \to D^* \pi$. The amplitude for a decay from the
helicity state $h$ to the $D^*$ state of helicity $k$ and a pion with
orientation  $(\theta, \phi)$ is proportional to
$$ Y_{2m}(\theta,\phi) \VEV{2 m 1 k\bigm| 2 h},
   \eqn\firstCleY$$
with $m = h-k$. Summing over $D^*$ helicities, and summing over $D_2^*$
helicities with the probabilities from \chelic, we find the following
result for the pion angular distribution:
$$ {1\over \Gamma}{d\Gamma\over d\cos\theta}(D_2^* \to D^* \pi) =
   \coeff38 \Bigl[ 1 + \cos^2\theta -  2 w_{3/2}
   \bigl( \cos^2\theta - \coeff 13\bigr)\Bigr] \ .
   \eqn\disttwo$$
This is a flatter distribution then we found for the direct decay to $D$.
\FIG\Dstards{Angular distribution of pions from the  decays $D_2^* \to
D \pi$ (solid), $D_2^* \to D^* \pi$ (dashed), and $D_1 \to D^*
\pi$. For $D_1$ decays, the dashed curve denotes the ideal case
of zero mixing (and is the same as for $D_2^*\to D^*\pi$), while the
dotted curve is computed for the more realistic situation $(S/D)^2 =
2$, $\eta=0.45$.  The curves assume the preferred value $w_{3/2}= 0$
and average over the polarization of the final $D^*$'s.} The two
distributions are compared in Fig.~\Dstards\ for the preferred value
$w_{3/2} = 0$.

Additional information can be obtained if the $D^*$ is observed through
its pion decay to $D$.  The amplitude for this secondary decay is
proportional to $Y_{1k}(\theta_2,\phi_2)$, where the angles give the
orientation of the secondary pion. The joint angular distribution of
the two pions is proportional to
$$ \sum_h  p(D_2^*,h)  \biggl| \sum_{k =-1,0,1}
   Y_{2m}(\theta,\phi) Y_{1k}(\theta_2,\phi_2)
   \VEV{2 m 1 k\bigm| 2 h}\biggr|^2,
   \eqn\secondYs$$
where, again, $m= h-k$.  In writing \secondYs, we ignore the $D^*$
recoil, as is appropriate in the heavy quark limit.  Simplifying this
expression, we find
$$ \eqalign{
   {1\over \Gamma}&{d\Gamma\over d\cos\theta d\cos\theta_2 d\phi_2}
   (D_2^* \to  \pi \pi  D) =\cr&
   {9\over 32\pi}
   \Bigl[ 1 + 2 \cos\theta \cos\theta_2 \cos\alpha - \cos^2\alpha
   - \cos^2 \theta_2 - \cos^2 \theta \cos^2 \alpha \cr
   &\qquad \qquad - 2w_{3/2}(\coeff13 +
   2 \cos\theta \cos\theta_2 \cos \alpha - \coeff 13 \cos^2
   \alpha -  \cos^2 \theta_2 - \cos^2\theta \cos^2 \alpha )\Bigr]\ ,\cr}
   \eqn\distthree$$
where
$$ \cos\alpha = \cos\theta \cos\theta_2 + \sin\theta \sin \theta_2
   \cos(\phi_2-\phi)
   \eqn\alphadef$$
is the angle between the two pions in the $D^*$ rest frame.  The
integral of this expression over $\theta_2,\phi_2$ reproduces \disttwo,
and the integral over orientations with $\alpha$ fixed gives the
$\sin^2\alpha$ distribution characteristic of the spin-2
parent.\refmark{\argus,\cleo}\ Notice that the complete distribution
\distthree\ is symmetric under $\cos\theta \to -\cos\theta$, so, again,
all information about the heavy quark polarization is lost.

The decay $D_1 \to D^* \pi$ can be analyzed in a similar fashion. In
the ideal   situation, we would ignore mixing of the $D_1$ with the
$D_1'$.  Then the decay amplitude  from $D_1$ helicity $h$ to $D^*$
helicity $k$ would be proportional to
$$ Y_{2m}(\theta,\phi) \VEV{2 m 1 k\bigm| 1 h} .
   \eqn\thirdYs$$
This would lead to a pion angular distribution
$$ {1\over \Gamma}{d\Gamma\over d\cos\theta}(D_1 \to D^* \pi) =
   \coeff38\Bigl[1  +  \cos^2\theta -  2 w_{3/2}
   \bigl( \cos^2\theta - \coeff 13\bigr)\Bigr] \ .
   \eqn\distfour$$
Curiously, this distribution is identical to \disttwo.

However, we have argued above that the $D_1$ must also have some
$S$-wave component to its decay due to mixing.  Following LWI, we
modify \thirdYs\  to
$$ Y_{2m}(\theta,\phi) \VEV{2 m 1 k\bigm| 1 h}
   - {S\over D}   e^{i\eta}\cdot
   Y_{00}(\theta,\phi) \delta(k,h)
   \eqn\thirdYs$$
The  parameter $S/D$ contains the $D_1$--$D_1'$ mixing angle and the
relative magnitudes of the  $D_1$ and $D_1'$ decay amplitudes. Note
that $S/D$ can be negative.  The phase $\eta$ of the interference term
is approximately equal to the the $D^*\pi$  $I = \coeff 12$ $S$-wave
phase shift; we do not call it $\delta_{1/2}$ to avoid confusion with
the Kronecker delta symbol $\delta(k,h)$. Extrapolating linearly from
the Weinberg value\Ref\WeinPS{S. Weinberg, \sl Phys. Rev. Lett. \bf 17,
\rm 616 (1966).} of the phase shift at threshold, we estimate
$$ \eta =  {1\over 4\pi} {m_\pi^2 \over f_\pi^2} \cdot
   {p_\pi\over m_\pi}  =  0.45
   \eqn\valofdelta$$
at the excitation energy of the $D_1$.  The inclusion of the $S$-wave
amplitude increases the width of the $D_1$ by a factor $(1 + (S/D)^2)$.
In our numerical examples, we will take $(S/D)^2 = 2$.  The inclusion
of the $S$-wave term dilutes the angular dependence of \distfour \ as
follows:
$$ \eqalign{
   {1\over \Gamma}{d\Gamma\over d\cos\theta}(D_1 \to D^* \pi)& =\cr
   {1\over 1 + (S/D)^2}\cdot {3\over 8}  \Bigl[&  1 + \cos^2\theta
   + \coeff 43 \bigl({S\over D}\bigr)^2
   - \coeff 23 \sqrt{2} {S\over D}\cos\eta (1-3 \cos^2\theta) \cr
   & \qquad -2w_{3/2}\bigl( \cos^2\theta -
   \coeff 13  - \coeff 23 \sqrt{2}{S\over D}\cos\eta
   (1-3\cos^2\theta) \bigr)\Bigr] \ . \cr}
   \eqn\distfive$$
The corrected pion angular distribution is compared to the idealized
form, and to our earlier results, in Fig.~\Dstards.

LWI suggested that the mixing parameter $S/D$ can be measured from the
properties of the joint pion angular distribution in $D_1 \to \pi D^*
\to \pi \pi D$.  They presented a number of useful partial
distributions.  But actually it is not difficult to construct the
complete joint distribution of the two pion momenta, since it is simply
proportional to
$$ \eqalign{
   &{1\over 1 + (S/D)^2} \times\cr
   &\quad\sum_h  p(D_1,h) \biggl| \sum_{k=-1.0.1}
   Y_{1k}(\theta_2,\phi_2) \big[ Y_{2m}(\theta,\phi)
   \VEV{2 m 1 k\bigm| 2 h}
   - {1\over \sqrt{ 4\pi}}{S\over D}
   e^{i\eta} \delta(k,h)\bigr]\biggr|^2\cr .}
   \eqn\fourthYs$$
The explicit formula for this angular distribution is:
$$ \eqalign{
   {1\over \Gamma}&{d\Gamma\over d\cos\theta d\cos\theta_2 d\phi_2}
   (D_1 \to  \pi \pi  D) =\cr&
   {1\over 32\pi}{1\over 1 + (S/D)^2}
   \Bigl[ 1 -18 \cos\theta \cos\theta_2 \cos\alpha +3 \cos^2\alpha
   +3\cos^2\theta_2 +27 \cos^2 \theta \cos^2 \alpha \cr
   & \qquad
   - 2w_{3/2}(-1 - 18\cos\theta \cos\theta_2 \cos \alpha
   - 3\cos^2\alpha + 3\cos^2\theta_2 +27\cos^2\theta \cos^2\alpha) \cr
   & \qquad +2 \bigl({S\over D}\bigr)^2
   \bigl(1 + 3\cos^2\theta_2 -2w_{3/2}(3\cos^2\theta_2 - 1)\bigr) \cr
   & \qquad-2\sqrt{2} {S\over D}\cos\eta \bigl( 1
   - 9\cos\theta \cos\theta_2 \cos\alpha -3 \cos^2\alpha +3\cos^2
   \theta_2 \cr
   &\qquad\qquad
   - 2w_{3/2}(-1 - 9\cos\theta \cos\theta_2 \cos \alpha
   +3 \cos^2\alpha + 3 \cos^2\theta_2  )\bigr)\cr
   &\qquad+ 6 \sqrt{2} {S\over D}\sin\eta \cos\alpha\cdot
   (1-4w_{3/2})\cdot(\hat 3\times\hat p_\pi\cdot \hat p_{\pi 2})
   \cdot P \Bigr] \ .\cr}
  \eqn\distthreenext$$
The invariant in the last line is the triple product of the
fragmentation axis with the directions of the two pion momenta.  We
have  multiplied this term by the original charmed quark polarization
$P$, since it is odd under reversal of the charmed quark spin
direction. The remaining terms in \distthreenext\ are independent of
$P$.  When the distribution \distthreenext  \ is integrated over angles
with $\alpha$ fixed, it gives a distribution intermediate between the
pure $D$-wave distribution $(1 + 3 \cos^2\alpha)$ and the flat
distribution expected from an $S$-wave decay.  Unfortunately, the
results on the $\alpha$ distribution reported in refs.~\argus\ and
\cleo\ are not yet sufficiently precise to give a useful constraint on
$(S/D)$.

The last term in \distthreenext\ is a counterexample to the no-win
theorem, the only one that we have found in the study of heavy meson
fragramentation.  It arises because the invariant
$$ \vec s \times \hat p_\pi \cdot \hat p_{\pi_2},
   \eqn\theinvariant$$
where $\vec s$ is the heavy quark spin, is parity-even and so can
appear in the angular distribution formula.\Ref\Nacht{O. Nachtmann,
          \sl Nucl. Phys. \bf B127, \rm 314 (1977).}
This invariant is
apparently $T$-odd, but this simply means that the contribution of the
invariant must be proportional to an absorptive phase.  In this case,
the phase is $\eta$, given approximately  by \valofdelta.  The phase is
sufficiently large that this effect might someday be used to confirm
that the $c$ quarks emerging from the $Z^0$ are predominantly
left-handed.

Since the $D_1$ and $D_2^*$ are prominent resonances of the charmed
mesons, it is natural that bottom mesons should possess similar excited
states.  We now briefly discuss the properties of those resonances.
The splitting of the heavy quark multiplet should decrease by a factor
$(m_b/m_c) \sim 3$ as we go from the charm to the bottom system, while
the decay rates remain roughly constant, up to angular momentum
factors.  Thus, we expect that the bottom mesons should have a set of
resonances located about 530 MeV above the centroid of the $(B,B^*)$
system.  These resonances should have widths of 20 MeV and a splitting
of 10 MeV.  The added width due to $B_1$--$B_1'$ mixing should be down
by a factor $(m_c/m_b)^2$ from the charm case; thus we can ignore this
effect here.   Note that the change to $b$ quarks interchanges the
relation of $\Gamma$ and $\Delta$ that we had for charm.

Since the bottom system has $\Gamma > \Delta$,  the two peaks
associated with the initial $B_1$ and $B_2^*$ should be merged.
However, since the $B$--$B^*$ splitting is 46 MeV, the separate decays
to $B$ and $B^*$ should be resolved.  Thus, we would expect that, when
$B$ mesons are produced in fragmentation, one should see two peaks in
the pion energy distribution in the $B$ meson frame, corresponding to
pion energies of about 520 and 565 MeV, each peak having a width of
about 20 MeV.  The relative populations of the two peaks should be 2:1
in favor of the lower-energy transition  $(B_1, B_2^*) \to B^*$; the
3:1 ratio from spin counting is partially balanced by a 1.5:1 ratio of
the kinematic factors $p_\pi^5$.  This experiment would allow both the
discovery of the $(B_1,B_2^*)$ multiplet and a nontrivial confirmation
of the $B$--$B^*$ mass splitting.

The fact that the $B_1$ and $B_2$ decay coherently has a curious effect
on the angular distribution of the decay pion.  In the limit $\Gamma\gg
\Delta$, we should compute this distribution as a decay of the $j^P =
\coeff32^+$ light antiquark configuration.  The angular distribution
for this decay is proportional to
$$ \sum_h p(\coeff 32,j^3) \biggl|
   Y_{2m}(\theta,\phi) \VEV{2 m \coeff 12 j^\pt\bigm| \coeff 32 j^3}
   \biggr|^2,
   \eqn\jthYs$$
where $p(\coeff 32, j^3)$ are the light antiquark probabilities from
\thethalfprobs \ and $j^\pt$ is the helicity of the light antiquark
after the decay.  Our formalism predicts that the two helicity states
$j^\pt = \pm \coeff 12$ are  equally populated; these populations then
can be combined with the heavy $b$ quark spin to form $B$ and $B^*$
mesons. For any $b$ polarization, the pion angular distribution follows
from \jthYs.  Working this out explicitly, we find
$$ {1\over \Gamma}{d\Gamma\over d\cos\theta}(B_1, B_2^* \to B,B^* \pi) =
   \coeff14\Bigl[1  + 3 \cos^2\theta - 6 w_{3/2}
   \bigl( \cos^2\theta - \coeff 13\bigr) \Bigr] \  ,
   \eqn\distjone$$
with the same distribution for the decay to $B$ and $B^*$.  This
distribution is identical to \distone, and that is easy to understand:
We can think of the decay amplitude to $B$ as a coherent sum of the
decay amplitudes from $B_1$ and $B_2^*$ to $B$; however, the amplitude
for $B_1 \to B \pi$ is  zero, and so we revert to the earlier case.
However, the relation of \distjone \ to \disttwo\ and \distfour\ is
quite surprising.  Naively, we might have expected the distribution in
this case to be an average of \disttwo\ and \distfour\ (which are
actually identical).  However, we find instead a sharper angular
distribution, as the result of the coherent superposition of the two
decay amplitudes.   The difference between \distjone\ and \disttwo,
\distfour\ reflects the loss of information on the spin of the light
degrees of freedom  which occurs when the heavy quark spin becomes
involved in the dynamics. By observing this transition from the charm
to the bottom system, we would effectively be timing the heavy quark
spin flip.

It should be noted that the calculation we have done applies to the
asymptotic case $\Gamma \gg \Delta$.  For $\Gamma$ and $\Delta$ of the
same order of magnitude, a more complicated formula is required.  We
give this formula in Appendix B.

We close this section with some speculations on the meaning of the
result $w_{3/2} =0$.  We have learned, in effect, that when a light
spin-$\frac32$ object forms in heavy quark fragmentation, its angular
momentum prefers to align transverse to, rather than along, the
fragmentation axis.  This is a striking result, and we have not been
able to find an explanation for it.  In models of string fragmentation,
the physical degrees of freedom of the string are transverse
oscillations, and so the orbital angular momentum would tend to point
along the string direction, that is, along the fragmentation axis.
Perturbative quark evolution by the Altarelli-Parisi equations can
produce correlations between  quark helicity and orbital angular
momentum.  For example, a polarized quark preferentially emits a gluon
with the same helicity and opposite orbital angular momentum. Some, but
not all, of this angular momentum can accompany an antiquark produced
from the gluon.  Neither viewpoint seems to lead to a crisp explanation
of the phenomenon.  In any event, this result on $w_{3/2}$, and related
results for other values of $w_j$ that will be found in the near
future, provide information on the process of fragmentation from a new
perspective.  Thus, they should provide incisive tests for proposed
schemes of hadronization.

\chapter{Polarization of Heavy Baryons}

We will now carry over the insights we have gained from the study of
heavy mesons to the phenomenology of heavy baryons.  For heavy mesons,
we saw that the `no-win' theorem prohibits any visible effects of an
initial heavy quark polarization, except under the special conditions
described below \theinvariant. However, for heavy baryons, the
situation is very different.  The ground state heavy baryon is built
from a heavy quark combined with a $j = 0$ combination of two light
quarks.   Since this system has no angular momentum to transfer to the
heavy quark, the initial polarization cannot be diluted. Mannel and
Schuler\refmark{\MS} and Close, K\"orner, Phillips, and
Summers\refmark{\CLKPS} have used this argument to conclude that the
ground state $b$ baryons  produced in $Z^0$ decays will retain the
initial high polarization $P$ of the $b$ quark. In this section, we
will compute the first correction to this argument and find the
depolarization of the $b$ quark in this scheme of fragmentation.   In
the process, we will explore the polarization dependence of excited
heavy baryon decays and find some further reactions which are sensitive
to the competition between the decay and the spin splitting  of a heavy
quark multiplet.

To begin, we review some basic properties of $b$ baryons.  Baryons are
expected about 5\%  of the time in $b$ fragmentation,\Ref\Saxon{D. Saxon,
in \sl High Energy Electron-Positron Physics, \rm A. Ali and
P. S\"oding, eds. (World Scientific, 1988).} so that about
10\% of $b\bar b$ events or 2\% of $Z^0$ hadronic events will contain
baryons.  In the nonrelativistic quark model, the lightest heavy
baryons consist of a heavy quark together with a light quark pair with
zero orbital angular momentum.  This pair can be either a $ud$ system
with isospin and spin $I = S = 0$ or a $uu$, $ud$, or $dd$ system with
$I= S= 1$.  (We ignore strange heavy baryons.) In the heavy quark
effective theory, the lightest baryons should be formed from states of
the light degrees of freedom with these quantum numbers. We will refer
to such states as `diquarks' even when we do not assume that the quark
model describes them accurately.   By combining the $s = \coeff 12$ $b$
quark with the diquarks of $j^P = 0^+$ and $1^+$, we form the
$\Lambda_b$ baryon and the $(\Sigma_b,\Sigma_b^*)$ baryon multiplet.
We will treat these three sets of states as the final states of the
rapid phase of $b$ fragmentation to baryons.

Even if we ignore the coupling of the $b$ quark spin, as is appropriate
to the  heavy quark approximation, the relative probabilities of
finding these states in $b$ fragmentation is still governed by two
unknown parameters.  The first of these, which we will call $A$, is the
relative probability of producing an $I=S=1$ diquark  as opposed to an
$I=S=0$ diquark. This is the ratio of the total $(\Sigma_b,\Sigma_b^*)$
production to primary $\Lambda_b$ production, summed over the 9
possible spin and isospin states of the $I=S=1$ multiplet.  The second
of these is the parameter $w_1$ which gives the probability that the
spin 1 diquark has maximum angular momentum $j^3 = \pm 1$ along the
fragmentation axis.  The parameter $A$ is related, but not identical,
to a parameter of the Lund fragmentation model  which gives the
relative probability of a spin 1 or a spin 0 diquark appearing when the
color string breaks:  $A \approx 9 \cdot {\rm PAR}(4)$.\Ref\Lund{T.
Sj\"ostrand, \sl Comp. Phys. Comm., \bf 39,  \rm 347  (1986).} An
important difference is that our parameter  $A$ is an output rather
than an input of the fragmentation scheme, so that it is defined
independently of any model. The parameter PAR(4) is not well determined
experimentally.  For example, in a recent study by the OPAL
collaboration,\Ref\OPAL{P. D. Acton, \etal\  (OPAL Collaboration), \sl
Phys. Lett. \bf B291, \rm 503  (1992).} this parameter could be varied
by a factor 3 from the Lund default value of 0.05 by adjusting the
other parameters of the baryon decay scheme. We know of no experimental
determination of $w_1$.  Nevertheless, it will be useful to have some
definite values of these parameters for our numerical estimates.
Motivated by the Lund default value and the results of the previous
section, we will choose
$$ A = 0.45 \ , \qquad w_1 = 0 \
   \eqn\defaults$$
as our reference values.  With these values, about 30\% of $b$ baryons
are born initially as $\Sigma_b$ or $\Sigma_b^*$.

We now consider the fragmentation of a $b$ quark with   complete
left-handed polarization. Given values of $A$ and $w_1$, the various
helicity states of the $b$ baryons are populated by fragmentation
according to the following table:
$$ \pmatrix{ p(\Sigma_b^*, h )\cr p(\Sigma_b,h)\cr p(\Lambda_b,h)\cr
   } \ = \ {1\over 1+A} \cdot
   \pmatrix{ \coeff12 w_1 A& \coeff23 (1-w_1) A& \coeff16 w_1A &
   \phantom{w}0\phantom{w}
   \cr & \coeff13(1-w_1)A & \coeff13 w_1 A & \cr   & 1 &  0 & \cr} .
   \eqn\Lbhelic$$
The probabilities for the $\Sigma_b$ and $\Sigma_b^*$ helicity states
represent the sum over the three isospin states.   The relative
production rate of $\Sigma_b : \Sigma_b^*$ is 1:2 independently of
$w_1$.

\REF\Lambdab{C. Albajar, \etal\ (UA1 Collaboration), \sl Phys. Lett. \bf
    B273, \rm 540 (1991);\break
    D. Decamp, \etal\ (ALEPH Collaboration), \sl Phys. Lett. \bf B278,
    \rm 209 (1992);\break
    P. D. Acton, \etal\ (OPAL Collaboration), \sl Phys. Lett. \bf B281,
    \rm 394 (1992);\break
    P. Abreu, \etal\ (DELPHI Collaboration), CERN-PPE/93-32 (1993).}
\REF\KRQ{W. Kwong, J. Rosner, and C. Quigg, \sl Ann. Rev.
    Nucl. Sci., \bf 37, \rm 325 (1987).}
\REF\ArgS{H. Albrecht, \etal\ (ARGUS Collaboration), \sl Phys.
    Lett. \bf B211, \rm 489 (1988);
    T. Bowcock, \etal\ (CLEO Collaboration), \sl Phys.
    Rev. Lett. \bf 62, \rm 1240 (1989);
    J. C. Anjos, \etal\ (Tagged Photon Collaboration), \sl Phys.
    Rev. Lett. \bf 62, \rm 1721 (1989).}
We next consider the mass splittings of the $b$ baryons. Unfortunately,
in the $b$ baryon system, only the $\Lambda_b$ is
known,\refmark\Lambdab\ and the only certain piece of information on
any heavy baryon splitting is: $m(\Sigma_c) - m(\Lambda_c) = 168$
MeV.\refmark{\PDG}    The $\Sigma_c^*$ has not yet been discovered. One
can estimate its position from the splittings of the strange baryons;
using quadratic mass relations, we find $m(\Sigma_c^*) - m(\Sigma_c) =
100$ MeV; for comparison, Kwong, Rosner, and Quigg\refmark\KRQ\ find 64
MeV for this mass difference using linear relations. The experiments
which give the $\Sigma_c$ mass\refmark{\ArgS} would seem to exclude
values of this mass difference below 80 MeV. Using our estimates, the
centroid of the $(\Sigma_c,\Sigma_c^*)$ multiplet is located 230 MeV
above the $\Lambda_c$.  The value of this mass splitting is expected to
have only a weak dependence on the heavy quark mass.  Thus, we expect
that the $\Sigma_b$ and $\Sigma_b^*$ should lie roughly 210 MeV and 240
MeV, respectively, above the $\Lambda_b$.  Both splittings are well
above the threshold for single-pion transitions to the $\Lambda_b$.
Thus, we expect that all $b$ baryon states will eventually decay
hadronically to $\Lambda_b$.

The decay rate for the transitions $(\Sigma_b,\Sigma_b^*) \to \pi +
\Lambda_b$ can be estimated in the nonrelativistic quark model by using
a pion-quark coupling estimated from the Goldberger-Trieman relation.
This computation has been done by Yan \etal\Ref\Yan{T.-M. Yan, H.-Y.
Cheng, C.-Y. Cheung, G.-L. Lin, Y. C. Lin, and H.-L. Yu, \sl Phys. Rev.
\bf D46, \rm 1148 (1992).} They find
$$ \Gamma = {g_{Aq}^2 \over  6\pi f_\pi^2} p_\pi^3  = 28\mev \cdot
  \bigl({ p_\pi\over 200\mev} \bigr)^3   ,
  \eqn\Yanest$$
where $p_\pi$ is the pion 3-momentum, $f_\pi = 93$ MeV, and $g_{Aq}$
is the axial vector coupling of the constituent quark.  In the
numerical estimate, we take  $g_{Aq} = 0.75$ to give the correct $g_A$
for the nucleon. The $\Sigma_b$ and $\Sigma_b^*$ have the same decay
rate up to kinematic factors, since the decay mechanism does not
directly involve the  heavy quark.

It is curious that the predicted decay rate $\Gamma$ and mass splitting
$\Delta$ for the $(\Sigma_b,\Sigma_b^*)$ multiplet are approximately
equal.  This is an accident, since $\Gamma$ is independent of the heavy
quark mass while $\Delta$ is proportional to $1/m_b$.  We have stressed
that our estimates of $\Delta$ and $\Gamma$ are quite uncertain.
However, if they are correct, the $\Sigma_b$ and $\Sigma_b^*$ form two
distinct resonances which thus decay incoherently.  The two excited
baryons can be observed together starting from a sample of (partially)
reconstructed $\Lambda_b$'s by plotting the distribution of pion
energies in the $\Lambda_b$ frame.  The $\Sigma_b$ and $\Sigma_b^*$
should appear as two closely spaced peaks on this distribution.  The
proper values of $\Gamma$ and $\Delta$ for the analysis to follow must
eventually be determined experimentally by the measurement of this
double-peak structure.

If it had turned out that $\Gamma \gg \Delta$, the $\Sigma_b$ and
$\Sigma_b^*$ baryons could decay to $\Lambda_b$'s without involving the
heavy quark spin.  In this limit, there would be no depolarization of
the $b$ quark from its initial polarization $P$.  However, our
estimates make it reasonable to consider the opposite limit in which
the two baryon resonances decay incoherently.  After we analyze this
limit in some detail, we will also present results for intermediate
values of $\Gamma/\Delta$.

Now we have all the ingredients we need to compute the properties of
the excited baryon decays and the effect of these decays on the
$\Lambda_b$ polarization.  We first consider the pion angular
distributions.  The amplitude for the decay of a $\Sigma_b$ of
helicity $h$ to a $\Lambda_b$ of helicity $k$ is proportional to
$$ Y_{1m}(\theta,\phi) \VEV{1 m \coeff12 k\bigm| \coeff12 h},
   \eqn\SigCleY$$
where $\theta$, $\phi$ give the pion orientation with respect to the
fragmentation axis and $m = h-k$.  The amplitude for $\Sigma_b^*$ decay
is given by the analogous formula with $j = \coeff 32$. Squaring and
summing with the probabilities from \Lbhelic, we find the pion angular
distributions
$$ \eqalign{
   {1\over \Gamma}{d \Gamma\over d \cos\theta}(\Sigma_b\to \Lambda_b
   \pi) &=   \coeff12 \cr
   {1\over \Gamma}{d \Gamma\over d \cos\theta}(\Sigma_b^*\to \Lambda_b
   \pi) &=   \coeff14\Bigl[ 1+3\cos^2\theta
   - \coeff92 w_1 \bigl( \cos^2\theta - \coeff13\bigr)
   \Bigr]  \ . \cr }
\eqn\distSigs$$
The first of these distributions is isotropic; the second becomes
isotropic at $w_1 = \coeff 23$.  This second distribution can be used
to determine $w_1$ experimentally.  For comparison, the pion angular
distribution in the case $\Gamma \gg \Delta$ is:
$$ {1\over \Gamma}{d \Gamma\over d \cos\theta}(\Sigma_b,\Sigma_b^*
   \to \Lambda_b \pi) =   \coeff32\Bigl[\cos^2\theta
   - \coeff32 w_1 \bigl( \cos^2\theta - \coeff13\bigr)\Bigr]  \ .
   \eqn\distSigsII$$
The intermediate situation can be analyzed using the formulae provided
in Appendix B.

On the other hand, we may integrate over the pion angles and look
instead at the distribution of final $\Lambda_b$ helicities which
result from a sample of completely left-handed polarized $b$ quarks.
Again, we consider the extreme limit $\Delta \gg \Gamma$. From
$\Sigma_b$ decay, we find
$$ { \Lambda_b( +\coeff12)\over\Lambda_b(-\coeff12)}
   =   {2-w_1\over 1 + w_1}  .
   \eqn\sigtolam$$
{}From $\Sigma_b^*$ decay, we find
$$ { \Lambda_b( +\coeff12)\over\Lambda_b(-\coeff12)}
   =   {2-w_1\over 4 + w_1}  .
   \eqn\sigstartolam$$
Summing over all primary and secondary $\Lambda_b$'s, we find
$$ { \Lambda_b( +\coeff12)\over \Lambda_b(-\coeff12) }
   =   {2(2-w_1)A \over 9 + A(5 + 2w_1)}  .
   \eqn\alltolam$$

To return to the situation of $Z^0$ decays, multiply the  corresponding
polarizations by the initial $b$ polarization $P$ given by \Porig.
Thus, we find  for the final $\Lambda_b$ polarization $P_\Lambda$ the
values
$$ P_\Lambda
   \ = \ \left\{ {1 + (1+4w_1)A/9\over 1 +  A }P\ , \
   {1 + w_1\over 3}P \ , \ - { 1 - 2w_1\over 3}P \right\},
   \eqn\Pprimevals$$
for $\Lambda_b$'s from the full sample, from $\Sigma_b^*$ decays, and
from $\Sigma_b$ decays, respectively.  Inserting the value from
\defaults, we find
$$ P_\Lambda
   \ = \ \big\{  0.72\ P \ , \ 0.33\ P \ , \ -0.33 P \big\}\ ;
   \eqn\PPrimeeval$$
with \Porig, this implies a 68\% polarization in the full sample of
$\Lambda_b$'s observed in $Z^0$ decay.  The minus sign in the last
entry of \PPrimeeval \ is not a misprint but rather a curious
prediction which would be very interesting to confirm.  We emphasize
again that these predictions are valid only if the $\Sigma_b$ and
$\Sigma_b^*$ are distinct resonances and revert to the  naive
prediction $P_\Lambda = 1 \cdot P$ in the limit where these resonances
completely overlap.

\FIG\Varypol{(a) Pion energy spectrum for decays $(\Sigma_b,\Sigma_b^*)
\to\Lambda_b+\pi$, for the case $\Gamma=\Delta=30$ MeV.  The upper
curve is the total spectrum, while the lower curve is the contribution
from the $\Lambda_b(-\coeff12)$ helicity state.  The spectrum is
computed using the formula for $d\Gamma/dE_\pi$ given in Appendix B.
 (b) The polarization
of the final $\Lambda_b$'s as a function of $\Gamma/\Delta$.  We show
the polarization of the full sample of $\Lambda_b$'s
as well as the separate contributions arising
from $\Sigma_b$ and $\Sigma_b^*$ decays.   These subsamples are defined
carefully in Appendix B.}
The intermediate case $\Gamma \sim \Delta$ can be treated by regarding
the $\Sigma_b$ and $\Sigma_b^*$ as partially overlapping resonances. We
present the formulae for this case in Appendix B.  In Fig.~\Varypol(a),
we show the pion energy spectrum for decays $(\Sigma_b, \Sigma_b^*)\to
\Lambda_b + \pi$, and the contributions to the spectrum from each
$\Lambda_b$ helicity state, for the case $\Gamma = \Delta = 30$ MeV. In
Fig.~\Varypol(b), we show how the three polarizations computed in
\PPrimeeval\ change as a function of the ratio $\Gamma/\Delta$.

Since the extreme limit $\Delta \gg \Gamma$ is well satisfied in the
case of charmed baryons, all of the results we have obtained in the
preceding paragraphs should also apply to the $\Lambda_c$, $\Sigma_c$,
$\Sigma_c^*$ system.  We predict a polarization of 48\% for
$\Lambda_c$'s produced in $Z^0$ decays. The parameter $w_1$ could well
be measured at CESR or in fixed target experiments, since the
distributions \distSigs\ are independent of the heavy quark
polarization.

We should, finally, comment on the measurement of the polarization of
$\Lambda_b$ baryons.  Close, K\"orner, Phillips, and
Summers\refmark\CLKPS\ and Amundson, Rosner, Worah, and
Wise\Ref\ARWW{J. F. Amundson, J. L. Rosner, M. Worah, and M. B. Wise,
\sl Phys. Rev. \bf D47, \rm 1260 (1993).}\ have proposed that the
absolute magnitude of the $\Lambda_b$ polarization can be obtained by
comparing the lepton distribution in semileptonic $b$ decays to the
spectator model, and the first set of authors have proposed additional
methods using the $\Lambda_b \to \psi \Lambda$ decay mode.  However, it
is important to note as well that the {\it relative} polarization of
two different samples of $\Lambda_b$'s can be obtained more easily by
observing any parity-violating forward-backward asymmetry with respect
to the fragmentation axis in $\Lambda_b$ decay. For example,   the
forward-backward asymmetry  of $\Lambda$ production in $\Lambda_b$
decays should be proportional to $P_\Lambda$ and can thus be used to
check the relative magnitudes of $P_\Lambda$ in the three samples
described in \Pprimevals.

\chapter{Conclusions}

In this paper, we have discussed a number of phenomena connected to
heavy hadron spectroscopy which are sensitive to the competition
between the rate of a hadronic decay and the rate of a heavy quark spin
flip.  We have seen that this competition can affect the angular
distributions observed for the decay of heavy hadrons and the degree of
polarization of heavy baryons.  Conversely, the properties of heavy
hadron decays can be used to measure a new set of fragmentation
parameters which we have called $w_j$, which provide nontrivial tests
of schemes of hadronization.

We have added two contributions to the study of the observability of
heavy quark polarization as viewed from the final state of the
hadronization process.  For heavy baryons, one expects a large
polarization; we have computed the leading effect of fragmentation
which degrades this polarization.  For heavy mesons, one generally
expects no observable polarization effects, though we have identified one
particular circumstance in which a polarization effect may be visible.

We look forward to further insights that will come from experiments on
the excited states of hadrons containing heavy quarks.

\APPENDIX{A}{A:  Isgur-Wise Theory of Hadronic Transitions Between
Heavy-Quark States}

In ref. \IW, Isgur and Wise presented the general theory of hadronic
transitions between states containing a single heavy quark. This theory
was presented in a telegraphic
 (Physical Review Letters) style, which somewhat concealed the
 elegant structure of  their formalism.     In this
appendix, we review their theory and supply a few additional formulae
which make this basic structure more clear.  We apply these formulae
in Sections 4 and 5 of this paper.

An excited state of a heavy hadron may decay to a lower-mass state
containing the same heavy quark by a strong interaction process in
which light hadrons are emitted.  In the examples of this paper, the
decay involves the emission of a single pion; however, the general
formalism depends only on the angular momentum of the emitted system.
To leading order as the heavy quark mass goes to infinity, the heavy
hadron does not recoil and the heavy quark does not flip its spin.
Thus, we have the following general structure:  The initial and final
states are composed of a heavy quark with spin $s = \coeff12$, combined
with light degrees of freedom of angular momentum $j$ for the initial
state and $j'$ for the final state to form heavy hadrons of
total spin $J$ and $J'$.  The transition from $j$ to $j'$ involves the
emission of a light hadronic system of angular momentum $L$ and does
not change the heavy quark spin.  These six angular momenta form a
tetrahedron, and so the rate of the process is governed by a Wigner
6-$j$ symbol.

More explicitly, we assign an invariant matrix element ${\bf M}$ as the
strength of the $j\to L +j'$ transition.  Then the decay rate from any
$J$ state in the $j + s$ heavy hadron multiplet is given by decomposing
the $J$ state in the $j+s$ basis, setting the decay rate of the $j$
state to be ${\bf M}$ times the appropriate  Clebsch-Gordon
coefficient, and then recombining the $j' +s$ states into the $J'$
appropriate to the final state.  Thus, the decay amplitude is given by
$$ \eqalign{
   {\cal A}((J^\p J^\pt\to J & J^3 + L m)  = \cr
   {\bf M}  & \cdot  \VEV{J^\p J^\pt \bigm| j^\p j^\pt s s^3}
   \VEV{L m j^\p j^\pt \bigm| j j^3}
   \VEV{j j^3 s s^3\bigm| J J^3} ,  \cr}
   \eqn\IWformula$$
summed over the intermediate values $s^3, j^3, j^\pt$. This is eq.~(1)
of ref.~\IW.  This expression can be rewritten the form\Ref\Edmonds{A.
R. Edmonds, \sl Angular Momentum in Quantum Mechanics. \rm (Princeton
University Press, 1957).}
$$ \eqalign{
   {\cal A}((J^\p& J^\pt\to J  J^3 + L m)  =\cr  {\bf M} &  \cdot
   (-1)^{L + j' + s + J}
   (2j+1)^{1/2} (2J'+1)^{1/2} \left\{\matrix{j' & j & L\cr
   J& J' & s\cr}\right\}  \VEV{L m J^\p J^\pt \bigm| J J^3} ,\cr}
   \eqn\IWformII$$
involving the Wigner 6-$j$ symbol.  The dependence on $J^\pt, m, J^3$
is given by the angular momentum Clebsch-Gordon coefficient for the
overall process, as must be so.

The formula \IWformII\ decouples the angular dependence of the hadronic
decay products from the  dependence of the decay amplitudes on the
position $J$ in the $j +s$ heavy quark multiplet.  Both aspects of this
equation are thus clarified.   The angular distribution of the decay
products is determined by the simple relation
$$ { \cal A}\ \sim \ \sum_m  Y_{Lm}(\Omega)
   \VEV{L m J^\p J^\pt \bigm| J J^3} ,
   \eqn\Clebschrel$$
for fixed $J^3, J^\pt$.  The total rate of hadronic decays from a state
$J$ in the $j + s$ multiplet depends on $J$ through the factor
$$ \sum_{J'} (2j+1)(2J' + 1)
   \left| \left\{\matrix{j' & j & L\cr
   J& J' & s\cr}\right\} \right|^2    = 1
   \eqn\Clebsum$$
by the standard orthogonality relation. Thus, the total decay rate is
independent of $J$, as predicted by the physical picture of Isgur and
Wise.

It is important to note, as Isgur and Wise do, that these relations
apply formally to the limit in which the heavy hadrons in each in the
$j+s$ multiplets are essentially degenerate.  In realistic situations,
there may be important corrections to these relations coming from
kinematic factors in the amplitude. For example, a decay which emits a
pion with angular momentum $L$ has a rate proportional to
$p_\pi^{2L+1}$.  This factor may vary significantly over the heavy
quark multiplet in cases of practical interest, for example, in the
$(D_1 , D_2^*) \to (D, D^*) + \pi$ transitions considered in Section 4.
In addition, the emission of high-energy pions may be suppressed by
form factors.  Isgur and Wise assume a suppression factor
$\exp[-p_\pi^2/(1 \gev)^2]$, but we omit this factor for simplicity.
It gives at most a 15\% correction to relative decay rates.  We
encourage the reader to keep this factor in mind, however, as
contributing to the  theoretical uncertainty of our heavy quark
predictions.

On the other hand, it is a major point of this paper that these
relations also do not apply  when the splitting within a $j+s$
multiplet  is much smaller than the hadronic widths of the heavy
hadrons.  The transition to this regime is discussed in Section 4.

\APPENDIX{B}{B:  Partial Coherence of Heavy Hadron Decays}

In this paper, we have mainly discussed heavy hadron decays in the
extreme limits $\Gamma \gg \Delta$ or $\Delta \gg \Gamma$. However, it
often happens that $\Gamma$ and $\Delta$ are of the same order of
magnitude, and so it is useful to have a formula which interpolates
between these two limits.  To obtain such a formula, we sum coherently
over the heavy hadron states $H$ and $H^*$ as distinct resonances.  In
the following discussion, we will use a language in which the decay
from $(H,H^*)$ procedes by emission of a single pion of angular
momentum $L$.  However, similar formulae apply to any strong
interaction decay.

We consider transitions from $H$ and $H^*$, of spin $J = j \pm \half$,
to a ground state hadron ${\cal H}$ of spin  $J'$. Let $E_\pi$ be the
pion energy and let  $E_J$ be the excitation energy of the resonance:
$E_J = m_H - m_{\cal H}$ for $H$, and similarly for $H^*$.  In the
heavy quark limit, $H$ and $H^*$ have the same width $\Gamma$.
Assume first that the light system which leads to $H$ and $H'$ has
angular momentum $(j,j^3)$ with respect to the fragmentation axis, and
that the heavy quark spin is initially polarized left-handed. Then the
amplitude for production of the state ${\cal H}$ in association with a
pion of energy $E_\pi$ in the angular momentum state $(L,m)$ is
$$ {\cal A}(j^3) =   \sum_J   \VEV{Lm J' J^\pt \bigm| J J^3}
   {A_J\over E_\pi - E_J + i \Gamma/2}  \VEV{j j^3 s -\coeff12 \bigm|
   J J^3}  \ ,
   \eqn\calAdefin$$
with $J^3 = j^3 - \coeff12$, $m = J^3 - J^\pt$.  The factor $A_J$ is
the prefactor of the Clebsch-Gordon coefficient in \IWformII. Only the
ratio of the two factors $A_J$ is important.  In the two examples
analyzed here,
$$ A_1\ :\  A_2    \ = \   1 \ : \ + \sqrt{3\over 5}
   \eqn\ratofampsD$$
for the $(D_1,D_2^*) \to D^* \pi$ transition, and
$$ A_{1/2}\ :  A_{3/2}  \ = \   1 \ : \ +1
   \eqn\ratofampsD$$
for the $(\Sigma_b,\Sigma_b^*) \to \Lambda_b \pi$ transition.

To find the dependence of the pion emission rate on $E_\pi$, we square
the amplitudes \calAdefin\ and sum them incoherently with the
probability distributions of the light degrees of freedom:
$$ { d\Gamma \over d E_\pi} \sim
   \sum_{j^3}   p(j,j^3)  \bigl|{\cal A}(j^3)\bigr|^2 .
   \eqn\thesumofjs$$
For $(D_1, D_2^*)$, we use \thethalfprobs; for $(\Sigma_b,\Sigma_b^*)$,
we use
$$ p(1,j^3) =   (\coeff12 w_1, (1-w_1), \coeff 12 w_1) \ .
   \eqn\theoneprobs$$

The resulting distribution of pion energies contains two overlapping
resonances; thus, there is some ambiguity in the assignment of observed
decays to one resonance or the other.  In constructing Fig.~\Varypol,
we have arbitrarily divided the distribution at the centroid of the
$(\Sigma_b,\Sigma_b^*)$ multiplet, $m_C =(m(\Sigma_b) +
2m(\Sigma_b^*))/3$. Pions with energy less than $m_C - m(\Lambda_b)$
were assigned to the $\Sigma_b$ sample; those with greater energy were
assigned to the $\Sigma_b^*$.

\ack

We are grateful to many friends who have aided this investigation. Our
work was originally stimulated by discussions with Charles Baltay. It
has profited from comments and criticism from Philip Burrows, Michael
Hildreth, Michael Luke, Sheldon Stone, Mark Wise, and our colleagues in
the SLAC theory group.

\endpage
\refout\endpage
\figout\endpage
\end